\documentclass[aps,epsf,twocolumn,showpacs,10pt,superscriptaddress]{revtex4-1}

\usepackage{graphicx,amsfonts,times,bm,amsmath,verbatim,color,array}

\newcommand{\e}{\varepsilon}

\usepackage{natbib}
\usepackage{graphicx,amssymb,amsmath,ifthen,braket,xcolor}
\usepackage[linktocpage=true,colorlinks=true,urlcolor=blue,linkcolor=red,citecolor=blue]{hyperref}
\usepackage{bm}

\begin{document}


\title{Chiral anomaly induced nonlinear Nernst and thermal Hall effects in Weyl semimetals}

\author{Chuanchang Zeng}
\thanks{Both authors SN and CZ contributed equally.}
\affiliation{Centre for Quantum Physics, Key Laboratory of Advanced Optoelectronic Quantum Architecture and Measurement(MOE),
School of Physics, Beijing Institute of Technology, Beijing, 100081, China}
\affiliation{Beijing Key Lab of Nanophotonics $\&$ Ultrafine Optoelectronic Systems,
School of Physics, Beijing Institute of Technology, Beijing, 100081, China}
\affiliation{Department of Physics and Astronomy, Clemson University, Clemson, SC 29634, USA}
\author{Snehasish Nandy}
\thanks{Both authors SN and CZ contributed equally.}

\affiliation{Department of Physics, University of Virginia, Charlottesville, VA 22904, USA} 
\author{Sumanta Tewari}
\affiliation{Department of Physics and Astronomy, Clemson University, Clemson, SC 29634, USA}

\date{\today}

\begin{abstract}
Chiral anomaly or Adler-Bell-Jackiw anomaly in Weyl semimetals (WSMs) has a significant impact on the electron transport behaviors, leading to remarkable longitudinal or planar electrical and thermoelectric transport phenomena in the presence of electromagnetic gauge fields. These phenomena are consequences of the imbalanced chiral charge and energy induced by chiral anomaly \textcolor{black}{in the presence of non-orthogonal external fileds, namely $\bm{E \cdot B}\neq 0$ or $ \bm{B \cdot \nabla }T \neq 0$ ($\bm{E, B}$, and $\bm{\nabla}T$ are the electric field, magnetic field, and thermal gradient respectively)}. We here propose another two fascinating transport properties, namely, the nonlinear planar Nernst effect and nonlinear planar thermal Hall effect induced by chiral anomaly in the presence of $\bm{B \cdot \nabla}T\neq 0$ in WSMs. Using the semiclassical Boltzmann transport theory, we derive the analytical expressions for the chiral anomaly induced nonlinear Nernst and thermal Hall transport coefficients and also evaluate the fundamental mathematical relations among them in the nonlinear regime. The formulas we find in this current work are consistent with that predicted for the nonlinear anomalous electrical and thermoelectric effects induced by Berry curvature dipole recently. Additionally, in contrast to the recent work, by utilizing the lattice Weyl Hamiltonian with intrinsic chiral chemical potential, we find that the chiral anomaly induced nonlinear planar effects can exist even for a pair of oppositely tilted or non-tilted Weyl cones in both time reversal and inversion broken WSMs. The chiral anomaly induced nonlinear planar effects predicted here along with the related parameter dependencies are hence possible to be realized in realistic WSMs in experiment.  

\end{abstract}

\maketitle

\section{Introduction}
Topological Weyl semimetals (WSMs) accommodating Weyl fermions have drawn tremendous attention due to their fascinating topological properties \cite{WSM_2011_Wen,WSM_2017_Binghai,WSM_2018_NP, TaAs_2015_Weng,TaAs_Huang_2015, MoTe2_2015_Yan,MoTe2_typeII_2015_Andrei}. 
In a WSM, the Weyl nodes, which are defined as the positions in momentum and energy space where non-degenerate bands linearly touch with each other, always appear in pairs with well-defined but opposite chiralities \cite{Nielson_1981_nogo}. Each node of the pairs individually act as the source ($+$ chirality) or sink ($-$ chirality) of the Berry curvature \cite{Niu_2010_berryphase}, which can be viewed as the effective magnetic field in momentum space. In the absence of external magnetic fields, many topological transport phenomena induced by the nontrivial Berry curvature have been discussed in the literature in the linear regime of WSMs. Examples include 
the anomalous Hall, Nernst and thermal Hall effects~\cite{AHE_2014Burkov,AHE_2014Fiete,AHE_2016Girish,AHE_2017po,AHE_2018Hess,AHE_2018Joseph, AHE_2017Balents, AHE_2018Kampert, ANE_2018Saha,ANE_2017Bardason,ANE_2020Li}, etc. Recently, it has been shown that the Berry curvature dipole (BCD), which generates nonlinear anomalous thermoelectric responses in time-reversal (TR) symmetric but inversion symmetry (IS) broken two-dimensional (2D) systems~\cite{Inti_2015_BCD,Inti_2018_BCD_notTMDCs, zzDu2019_NLAHE_1,Law2019_NLAHE_3,qMa2019_NLAHE_experiment,kKang2019_NLAHE_experiment,Lee_NLAHE_4,Nandy_sysmetris_2019,Yu_NLANE_1,Zeng_NLANE_2,zeng2019wiedemannfranz}, can also manifest itself in three-dimensional (3D) WSMs and shows interesting electrical and optical effects in the nonlinear response regime \cite{Zhang_2018_BCD_abinitio, AHE_weyl2018Marco,AHE_weyl2020czeng}. 

A remarkable topological aspect in WSMs is the chiral anomaly \cite{CA_1969jb,CA_1969jb,CA_2012Aji,CA_2013Spivak}, which, in the presence of parallel electromagnetic fields ($\bm{E\cdot B \neq 0}$), leads to the non-conservation of the chiral charge. The chiral anomaly in WSMs gives rise to several intriguing magnetotransport phenomena, including the negative longitudinal magnetoresistance \cite{CA_2013Spivak,NM_2013Qi,NM_2014Burkov,NM_2014Sasaki,NM_2015prx,NM_CA_2016npONG}
and the planar Hall effect \cite{PHE_2017Burkov,PHE_2017Tewari,PHE_2018Mandal,PHE_2018Shekhar,phe_2018Sun,PHE_2019Zhang,PHE_2019ZhangAIP, NM_2015cHEN}, which have been well studied in both theories and experiments.
Nevertheless, their counterparts in the nonlinear regime remain barely discussed so far.
It has recently been proposed that the nonlinear planar Hall effect (NPHE) induced by the chiral anomaly can exist in tilted Weyl semimetals, originating from the combined effect of the Berry curvature related anomalous velocity and the modified carrier density induced by the chiral anomaly \cite{chiralanomaly_2020Li, Nandy_multiWSM_CA2021}. The chiral anomaly in this newly proposed nonlinear Hall effect belongs to the conventional electrical chiral anomaly, which requires the simultaneous presence of a non-orthogonal electric and magnetic fields. 

\textcolor{black}{Analogously, the presence of co-planar thermal gradient and magnetic field, i.e. $\bm{B\cdot \nabla}T \neq 0$, can also result in the chiral charge pumping and the non-conservation of the energy densities in WSMs. This is the so-called thermal chiral anomaly}. Not only the thermal chiral anomaly generated magnetotransport behaviors have been theoretically studied~\cite{thermalCN_2020das} in Weyl materials but also a giant enhancement on thermal conductivity induced by thermal chiral anomaly has been observed experimentally in topological bismuth-antimony alloys (Bi$_{1-x}$Sb$_x$)~\cite{thermalCN_2020Vu}. However, both of these works fall within the linear response regime. Here in this work, we propose another two fascinating planar thermoelectric effects in the nonlinear response regime that are induced by the thermal chiral anomaly, namely, nonlinear planar Nernst effect (NPNE) and nonlinear planar thermal Hall effect (NPTHE). To the best of our knowledge, these two chiral anomaly induced nonlinear planar effects proposed in this work, have not yet been discussed and can be probed experimentally. 

In this work, using the semiclassical Boltzmann transport approach with a relaxation time approximation, we derive the general expressions for the nonlinear planar Nernst and planar thermal Hall effect induced by the thermal chiral anomaly (Eqs.~(7), (9)). By doing Sommerfeld expansion in the low-temperature regime, we also obtain the fundamental relations among the chiral anomaly induced nonlinear planar thermoelectric transport coefficients. We find that in the nonlinear response regime ($\propto (\bm{\nabla} T)^2$), the chiral anomaly induced NPNE and NPHE coefficients are directly proportional to each other while NPTHE and NPHE coefficients are connected by a derivative relationship (Eq.~(13)). The fundamental formulas derived in this work 
remarkably reproduce relations identical to those predicted recently for the BCD induced nonlinear anomalous thermoelectric transport coefficients in the absence of magnetic field \cite{zeng2019wiedemannfranz}.

Using the general expressions of the nonlinear planar thermoelectric effects, we predict the behavior of these transport coefficients in both TR and IS broken WSMs using low energy linearized Weyl Hamiltonian as well as the lattice Weyl Hamiltonian. We consistently find that at chemical potentials away from the Weyl nodes, the chiral anomaly induced NPNE and NPTHE coefficients are proportional to $(k_B T)^0 \mu^{-2}$ and $(k_BT)^2 \mu^{-3}$ respectively, agreeing well with the nonlinear analog of the Wiedemann-Franz law and Mott relation derived in this work. 
Specifically, based on our numerical calculations via the lattice Hamiltonian with finite intrinsic chiral chemical potential, we find that the nonlinear planar transport coefficients can be non-zero even when the Weyl nodes are oppositely tilted or not tilted at all, in contrast to what has been found in a recent work \cite{chiralanomaly_2020Li}. The finite chiral chemical potential along with the lattice regularization naturally generate an asymmetric Fermi surface near the Weyl nodes in WSMs, resulting in non-zero net value of the Fermi surface contributions. Consequently, the results obtained via lattice model in this work, can be taken as a justifiable theoretical prediction in favor of the feasibility of probing of the nonlinear planar effects induced by chiral anomaly in realistic materials. 
The behavior of the newly proposed chiral anomaly induced NPNE and NPTHE effects predicted in this work can also be directly tested in experiments, e.g. via the frequency lock-in
measurement under frequency-dependent thermal gradient, in the TR symmetry broken as well as IS broken Weyl systems. 

The rest of the paper is organized as follows: In Sec.~II, we start with the semiclassical Boltzmann transport formalism, and in part A, we derive the expressions for the chiral anomaly induced nonlinear planar transport charge and heat currents for the configuration $\bm{E=0}, \bm{B\cdot \nabla}T \neq 0$; then we derive the fundamental relations among the chiral anomaly induced nonlinear planar transport coefficients via Sommerfeld expansion in part B. In Sec. III, we apply our analytically derived equations to the linearized Weyl Hamiltonian as well as the lattice Weyl Hamiltonian, and numerically check the parameter dependencies for the nonlinear planar Nernst and thermal Hall effects. Finally we end with a brief summary and conclusion in Sec. IV.



\section{Semiclassical Boltzmann transport formalism modified by Berry curvature }
The dynamics of the non-equilibrium distribution function $f({\bm{k,r},t})$ for Bloch electrons is phenomenologically described by the following Boltzmann transport equation, 
\begin{equation}
    \bigg(\frac{\partial}{\partial t} + \bm{\dot{r} \cdot \nabla_r } + \bm{\dot{k} \cdot \nabla_k}\bigg) f(\bm{r, k},t) = \it{I_{coll}}\big[ f(\bm{r, k},t) \big]
\end{equation}
where the right hand side represents a collision term which incorporates the effects of electron interaction and impurity scattering. The Berry curvature effect can be introduced into the above Boltzmann equation through the semiclassical equations of motion for the carriers \cite{moe_Berry_2006PC,moe_Berry_2012Yamamoto,Niu_2010_berryphase}, given as below,
\begin{equation}\begin{split}
 \bm{\dot{r}} & = D\big[\bm{v}+\frac{e}{\hbar}\bm{E \times \Omega} + \frac{e}{\hbar}(\bm{v \cdot \Omega })\bm{B}\big] \\
 \bm{\dot{k}} & =D \big[ -\frac{e}{\hbar}\bm{E} -\frac{e}{\hbar} \bm{v \times B} -\frac{e^2}{\hbar^2}(\bm{E \cdot B}) \bm{\Omega}\big]
\end{split}
\end{equation}
where $D$ is the shorthand for $D(\bm{B, \Omega}) =(1 + e (\bm{B \cdot \Omega})/\hbar)^{-1}$, the phase volume factor revealed in the presence of non-zero Berry curvature $\bm{\Omega}$ and magnetic field $\bm{B}$. Here, $\bm{v}=\hbar^{-1} \partial \e_{\bm{k}}/{\partial \bm{k}}$ is the carrier group velocity coming from the band dispersion $\e_{\bm{k}}$. The second and third terms in the $\bm{\dot{r}}$ equation give rise to the anomalous Hall effect 
and the chiral magnetic effect respectively, while the third term in the $\bm{\dot{k}}$ equation proportional to $(\bm{E\cdot B})$ is the source of chiral anomaly \textcolor{black}{in Dirac and Weyl semimetals}. 

In this work, we are interested in the steady-state solutions to the Boltzmann transport equation in a configuration of $\bm{E}=0$ but non-zero magnetic field $\bm{B}$ and thermal gradient $-\bm{\nabla}T$. Plugging Eq.~(2) into Eq.~(1), the Boltzmann transport equation can be rewritten as, 
\begin{equation}
    \begin{split}
        D\bigg[ \bm{v}+\frac{e}{\hbar} (\bm{ v\cdot \Omega})\bm{B}\bigg] \bm{\nabla_r}f_{\bm{r, k}} =-\frac{f_{\bm{r, k}}-f_{eq}}{\tau}
    \end{split}
\end{equation}
where we have invoked the relaxation time approximation with $\tau$ being the phenomenal scattering time. \textcolor{black}{We want to stress that, a more complete analysis should involve both the inter-node and intra-node scattering times for WSMs in the above equation. However, the leading contribution to the nonlinear responses in WSMs studied in this work can be found to arise from the inter-node scattering relaxation (see the Appendix). For simplicity, we here consider a single effective scattering time as applied in the recent works~\cite{AHE_2014Fiete,tau_intranode_2019Amit}, which will not qualitatively affect our results and conclusions in WSMs as shown in following sections.}

The equilibrium Fermi-Dirac distribution function is given as $f_{eq}=1/(e^{\beta(\e_{\bm{k}}-\mu})+1)$, for which at zero temperature we have $f_{eq} = \Theta(\mu-\e_{\bm{k}})$, and $\partial f_{eq}/\partial \e_{\bm{k}}=-\delta(\mu-\e_{\bm{k}})$ ($=-\delta(k-k_F)/|\partial \e_{\bm{k}}/\partial k|$).
To obtain non-linear thermoelectric responses second order in thermal gradient, we are interested in the correction terms $\delta f_{\bm{k}}=(f_{\bm{r, k}}-f_{eq})$ which can be formulated as $\delta f_{\bm{k}} =\sum_n f^n_{\bm{k}}$ with $f^n_{\bm{k}} \propto (\bm{\nabla}T)^n$. The first two correction terms are found as below, 
\begin{equation}
    \begin{split}
       & f^1_{\bm{k}} = \tau D \frac{\e_{\bm{k}}-\mu}{T} \big(\bm{v}+\frac{e}{\hbar}(\bm{v \cdot \Omega})\bm{B}\big)\bm{\nabla}T \frac{\partial f_{eq}}{\partial \e_{\bm{k}}}\\
       & f^2_{\bm{k}} = \tau D \frac{\e_{\bm{k}}-\mu}{T} \big(\bm{v}+\frac{e}{\hbar}(\bm{v \cdot \Omega})\bm{B}\big)\bm{\nabla}T \frac{\partial f^1_{\bm{k}}}{\partial \e_{\bm{k}}}
    \end{split}
\end{equation}
Note that, $f^1_{\bm{k}} \propto \tau$ while $f^2_{\bm{k}} \propto \tau^2$. \textcolor{black}{We should mention, that the orbital magnetic moment is ignored here in this work for simplicity, as its effect on the chiral anomaly induced nonlinear responses is found to be negligible in recent works
~\cite{chiralanomaly_2020Li, Nandy_multiWSM_CA2021}}. With the help of the above equations, we will derive the equations for the chiral anomaly induced nonlinear planar Nernst effect and nonlinear planar thermal Hall effect in what follows.  

\subsection{Chiral anomaly induced nonlinear planar Nernst effect and nonlinear planar thermal Hall effect}
Using the semiclassical equations of motion for the electrons, the charge current can be written as \cite{Nernstcurrent_2006Niu,Niu_2010_berryphase}, 
\begin{equation}
    \begin{split}
        \bm{j} &= -e\int [d\bm{k}] \bm{\dot{r}} f_{\bm{k}} +\frac{e k_B \bm{\nabla} T}{\hbar} \times \int [d\bm{k}]  \bm{\Omega_k}s_{\bm{k}} \\
    \end{split}
\end{equation}
where $s_{\bm{k}}=-f_{eq}\mathbf{log}f_{eq}-(1-f_{eq}) \mathbf{log}(1-f_{eq})$ is the entropy density. In a more generalized format, the non-equilibrium distribution function $f_{\bm{k}}$ can be applied to $s_{\bm{k}}$
to generate the higher order (over linear order) responses \cite{Yu_NLANE_1, Zeng_NLANE_2}. Therefore, in the presence of thermal gradient and magnetic field, the generalized charge current can be rewritten as, 
\begin{equation}
    \begin{split}
        \bm{j}= & -e\int [d\bm{k}] D^{-1} \big[ \bm{v}+\frac{e}{\hbar}(\bm{v\cdot \Omega})\bm{B}\big] f_{\bm{k}}
        -\frac{\bm{e \nabla}T}{\hbar T} \bm{\times} \\ & \int [d\bm{k}]  \bm{\Omega}\bigg[(\e_{\bm{k}}-\mu)f_{\bm{k}} + \beta^{-1} \mathbf{log}(1+e^{-\beta(\e_{\bm{k}}-\mu)})\bigg]
    \end{split}
\end{equation}
A linear planar thermopower contribution ($\propto (\bm{\nabla}T)$) can be extracted from the first term in the above equation, as has been studied recently for Dirac and Weyl semimetals \cite{linearthermopower_2019Tewari}. The second term in the above equation, on the hand, describes the purely anomalous Nernst effect independent of magnetic field in the linear response regime. However, the second term can support nonlinear effects when the higher order correction terms of $f_{\bm{k}}$ are considered. Here in this paper, we are interested in the higher order responses induced by thermal chiral anomaly (i.e., the coplanar thermal gradient and magnetic field) via the contributions coming from the perturbed correction terms of the distribution function for carriers ($\delta f_{\bm{k}}$). After some algebra with Eq.~(6), the second order ($\propto (\bm{\nabla} T)^2$) planar Nernst response induced by chiral anomaly can be expressed as, 
\begin{equation}
    \begin{split}
        \bm{j}^N = \frac{e^2 \tau}{\hbar^2} & \int [d\bm{k}] \frac{(\e_{\bm{k}}-\mu)^2}{T^2} \frac{\partial f_{eq}}{\partial \e_{\bm{k}}} (\bm{\nabla}T \bm{\times \Omega})\\
        & \bigg[ (\bm{v\cdot \Omega})(\bm{B \cdot \nabla }T)
       - (\bm{B\cdot \Omega})(\bm{v \cdot \nabla}T)
        \bigg]
    \end{split}
\end{equation}
where the superscript `$N$' implies for the nonlinear contribution. Note that, the above equation is $\tau-$dependent, while in expanding Eq.~(6) with the help of Eq.~(4), we can get additional second order responses proportional to $\tau^2$ or $\bm{B}^2$. These additional terms, which in principle can be distinguished from the above contribution in experiment by their different scalings \textcolor{black}{in either $\tau$ or $B$}, are not the interest of this work and are ignored here. As shown by the above equation, the charge current vanishes either if $\bm{\nabla}T=0$ or $\bm{B}=0$. The first term ($\propto (\bm{B\cdot \nabla}T)$) inside the square bracket in Eq.~(7) is purely induced by chiral anomaly, and 
an \textit{effective} chiral anomaly induced contribution to the nonlinear planar Nernst current can be obtained from the second term. It is worth noting that, there is also an effective contribution proportional to $(\bm{B \times \nabla}T)$ contained in the second term inside the square bracket in Eq.~(7) to the nonlinear Nernst current, which identically vanishes in the configuration of $(\bm{B  || \nabla}T)$ and is not the interest of this work.  

Analogously, we should also expect the existence of nonlinear planar thermal Hall effect induced by the thermal chiral anomaly in response to the coplanar thermal gradient and magnetic field ($\bm{B \cdot\nabla}T$). 
In the presence $-\bm{\nabla}T$, the total transport thermal Hall current in a generalized format is given by \cite{THEcurrent_2010dORON}, 
\begin{equation}
    \begin{split}
        \bm{j}^Q_T = & - \frac{k^2_B T}{\hbar} \bm{\nabla}T \bm{\times} \int [d\bm{k}]  \bm{\Omega_{\bm{k}}}  \bigg[  \beta^2 \left(\e_{\bm{k}}-\mu \right)^2 f_{\bm{k}} + \frac{\pi^2}{3} \\ 
 & -\operatorname{In}^2\left(1-f_{\bm{k}}\right)- 2\operatorname{Li}_2\left(1-f_{\bm{k}}\right)\bigg]
    \end{split}
\end{equation}
Here, the superscript `$Q$' and subscript `$T$' represent for `heat current' and `thermal contribution' respectively. Following a similar analogy as discussed above for the nonlinear planar Nernst effect (Eq.~(6)-(7)), the chiral anomaly induced planar thermal Hall current in second order of thermal gradient ($\propto (\bm{\nabla}T)^2$) is found as, 
\begin{equation}
    \begin{split}
         \bm{j}^{Q,N}_T
           =  -\frac{e \tau T }{\hbar^2}  & \int [d\bm{k}]   \frac{\left(\e_{\bm{k}}-\mu \right)^3}{T^3} \frac{\partial f_{eq}}{\partial \e_{\bm{k}}} (\bm{\nabla}T \bm{\times} \bm{\Omega_{\bm{k}}})  \\
           & \bigg[ (\bm{v\cdot \Omega})(\bm{B \cdot \nabla}T)
          - (\bm{v\cdot \nabla}T) (\bm{B \cdot \Omega}) \bigg] \\
    \end{split}
\end{equation}
where we focus on the second-order nonlinear response which is linear in $\tau$, quadratic in thermal gradient and linear in magnetic field. Note that, we retain only the leading contributions to $\bm{j}^{Q, N}_T$ in Eq.~(9), where the other correction terms of orders of magnitudes much smaller than the leading term are omitted (valid in the limit of $\mu \gg k_B T$) \cite{zeng2019wiedemannfranz}. Both the nonlinear planar Nernst (Eq.~(7)) and thermal Hall effect (Eq.~(9)) are dependent on the derivative of the Fermi distribution function with respect to energy ($\partial f_{\bm{k}}/\partial \e_{\bm{k}}$), rendering these two nonlinear effects induced by chiral anomaly being Fermi surface quantities, similar as the BCD induced nonlinear effects \cite{Inti_2015_BCD,Yu_NLANE_1,zeng2019wiedemannfranz}. 
The chiral anomaly induced nonlinear currents $\bm{j}^N$ and $\bm{j}^{Q, N}_{T}$ can not survive under the presence of either time reversal symmetry or inversion symmetry, due to the symmetry properties of quantities $\bm{\Omega}_{k}, v_{\bm{k}},\e_{\bm{k}}$ in Eqs.~(7), (9).

Note that, apart from a different weight component $(\e_{k}-\mu)^3/T^3$, the nonlinear planar thermal Hall current described above appears in a similar format as that of the nonlinear planar Nernst current in Eq.~(7).  This fact allows us to acquire their remarkable connections to the nonlinear planar Hall coefficient induced by chiral anomaly by Sommerfeld expansion in the low temperature regime, which will be soon discussed in the following part. 

\subsection{Nonlinear analog of Wiedemann-Franz law and Mott relation in the presence of magnetic field }

In the regime of linear response, the fundamental relations among the thermoelectric transport coefficients (i.e., the  coefficients of electric Hall effect, Nernst effect and thermal Hall effect) are encapsulated by the well-known phenomenological Wiedemann-Franz law and Mott relations \cite{Lorenz_law, Ashcroft_my}. Very recently, the nonlinear analog of these two fundamental relations have been found in totally different forms for the BCD induced nonlinear anomalous transport phenomena in the time reversal symmetric systems \cite{zeng2019wiedemannfranz}. However, the relations among the planar transport coefficients induced by chiral anomaly remain unknown.

Here in this work, we focus on the configuration $\bm{\nabla}T =\nabla_x T \hat{\bm{x}}, \bm{B}=B_x \hat{\bm{x}}+B_y \hat{\bm{y}}$, namely the thermal gradient and magnetic field lying within the $xy-$plane and forming an angle $\theta$ such that $\bm{B\cdot \nabla}T = B {\nabla}_x T \cos{\theta}$. \textcolor{black}{As has been discussed in previous section, both the terms proportional to $(\bm{B \cdot \nabla} T)$ and $(\bm{B \times \nabla} T)$ can be obtained through Eqs.~(7) and (9).}
Within the configuration of $\bm{\nabla} T$ and $\bm{B}$ given above, we can extract the currents coming from the (thermal) chiral anomaly as being proportional to $\propto B \nabla_x T \cos{\theta}$. Without losing any generality, we will focus on the chiral anomaly induced nonlinear planar currents along $y-$direction in this present setup, which can be given by $\bm{j}^{CN}_{yxx} = \alpha_{yxx} (\nabla_x T)^2$ and $ \bm{j}^{Q,CN}_{yxx} = l_{yxx} (\nabla_x T)^2$ (superscript `CN' indicates the chiral anomaly induced nonlinear contribution). The conductivity $\alpha_{yxx}, l_{yxx}$ are then respectively calculated as
\begin{equation}
    \begin{split}
    \alpha_{yxx} & =c_{\alpha}\int [d\bm{k}] \frac{\partial f_{eq}}{\partial \e_{\bm{k}}} \textcolor{black}{\widetilde{\bm{\Omega}}^{\alpha}_{\bm{k}}} B\cos{\theta} \\
    l_{yxx} &= c_{l}\int [d\bm{k}] \frac{\partial f_{eq}}{\partial \e_{\bm{k}}}\textcolor{black}{\widetilde{\bm{\Omega}}^{l}_{\bm{k}}} B\cos{\theta}
    \end{split}
\end{equation}
where the coefficient  $c_{\alpha}= \frac{e^2  k^2_B \tau}{\hbar^2}, c_{l}= -\frac{e  k^2_B \tau }{\hbar^2}$ and the corresponding modulated Berry curvatures \textcolor{black}{$\widetilde{\bm{\Omega}}^{\alpha}_{\bm{k}}$ and $ \widetilde{\bm{\Omega}}^{l}_{\bm{k}}$} are respectively defined as 
\begin{equation}
    \begin{split}
    \textcolor{black}{\widetilde{\bm{\Omega}}^{\alpha}_{\bm{k}}}  & = \frac{(\e_{\bm{k}}-\mu)^2}{k^2_B T^2} \Omega^z_{\bm{k}} \bigg[\bm{v_k\cdot \Omega_k}-v_x\Omega^x_{\bm{k}}\bigg]  \\
   \textcolor{black}{\widetilde{\bm{\Omega}}^{l}_{\bm{k}}} &= \frac{(\e_{\bm{k}}-\mu)^3}{k^2_B T^2} \Omega^z_{\bm{k}} \bigg[\bm{v_k\cdot \Omega_k}-v_x\Omega^x_{\bm{k}}\bigg]
    \end{split}
\end{equation}
\textcolor{black}{Here the subscripts in $c_{\alpha, l}$ and the superscripts in $\Omega^{\alpha, l}_{\bm{k}}$ imply the quantities are related to nonlinear Nernst effect ($\alpha$) or nonlinear thermal Hall effect ($l$), respectively.} For the sake of simplicity, we have suppressed the index for the magnetic field component in $\alpha_{yxx}, l_{yxx}$ in Eq.~(10), given that we are only considering the component of the magnetic field parallel to thermal gradient. Note that, the aforementioned linear thermopower (Ref.~\cite{linearthermopower_2019Tewari}) that may arise from the first term in Eq.~(6), along with the other possible linear magnetotransport behaviors induced by the thermal chiral anomaly (Ref.~\cite{thermalCN_2020das}) in this current scenario, explicitly show different angular and field dependencies from that of the chiral anomaly induced nonlinear transport coefficients described by Eq.~(10). Hence, these features can in principle be used to distinguish the chiral anomaly induced nonlinear planar effects discussed here from other similar linear or nonlinear effects in experiments. 

To analyze the relationship among the transport coefficients of the chiral anomaly induced nonlinear planar transport phenomena (namely, the nonlinear planar Nernst and thermal Hall effects, and the nonlinear planar Hall effect), let us first recall the chiral anomaly induced nonlinear planar Hall conductivity \cite{chiralanomaly_2020Li}. Following a similar analogy, the chiral anomaly induced nonlinear planar Hall conductivity $\sigma_{yxx}$ can be written as,
\begin{equation}
    \begin{split}
  \sigma_{yxx} &=  c_{\sigma}\int [d\bm{k}] \frac{\partial f_{eq}}{\partial \e_{\bm{k}}}  \widetilde{\bm{\Omega}}_{\bm{k}, \sigma} B \cos{\theta}
    \end{split}
\end{equation}
where $c_{\sigma}=\frac{e^4 \tau}{\hbar^3}$ and the corresponding modulated Berry curvature $\widetilde{\bm{\Omega}}_{\bm{k}, \sigma}$ is defined as $ \widetilde{\bm{\Omega}}_{\bm{k}, \sigma}=\Omega^z_{\bm{k}} \big[\bm{v_k \cdot \Omega_{k}}-v_x \Omega^x_{\bm{k}}\big]$. Here a configuration with $\bm{E}=E\hat{\bm{x}}, \bm{B}=B_x \hat{\bm{x}}+B_y \hat{\bm{y}}$ for the chiral anomaly induced nonlinear planar Hall effect has been considered. 

In light of the results given by Eqs~(10)$-$(12), it is straightforward to derive the relations among the three chiral anomaly induced thermoelectric transport coefficients using the Sommerfeld expansion in low temperature regime \cite{Ashcroft_my, zeng2019wiedemannfranz}. By considering only the leading order terms for the expansions in terms of temperature, we find
\begin{equation}
    \begin{split}
        \alpha_{yxx} &=\frac{\pi^2 k^2_B}{3 e^2} \sigma^0_{yxx} + \mathcal{O}(T^2) \\
        l_{yxx} & = -\frac{7 \pi^4 k^4_B T^2}{15 e^2} \frac{\partial \sigma^0_{yxx}}{\partial \mu} +\mathcal{O}(T^4)
    \end{split}
\end{equation}
where $\sigma^0_{yxx}$ is the chiral anomaly induced nonlinear planar Hall conductivity at zero temperature (in Eq.~(12)). Different from the conventional Wiedemann-Franz law and Mott formula (i.e., $\kappa =L T \sigma$  and $\alpha = eLT \partial \sigma/\partial \mu$, respectively, with $L$ being the Lorentz number) in the linear regime \cite{Lorenz_law,Ashcroft_my}, the role of the \textit{derivative} with respect to chemical potential is now interchanged in the above relations describing the chiral anomaly induced nonlinear responses. To be more specific, in contrast to the linear regime, where the Nernst coefficient is proportional to the \textit{derivative} of the Hall coefficient along with a $T$-dependent proportionality factor ($\alpha = eLT \partial \sigma/\partial \mu$), in the nonlinear regime, these two chiral anomaly induced nonlinear transport coefficients are directly proportional to each other and the corresponding proportionality factor is $T$-independent (first line in Eq.~(13)). On the other hand, unlike the Wiedemann-Franz law describing the linear responses where the thermal Hall and charge Hall coefficients are proportional to each other ($\kappa =L T \sigma$), 
the counterpart of this relation in the nonlinear regime shows that the thermal Hall coefficient is proportional to the \textit{first-order derivative} of the Hall coefficient with respect to chemical potential (second line in Eq.~(13)). Additionally, in contrast to the regime of linear response, the proportionality factor is $T^2$- instead of $T$-dependent in the nonlinear analog of the Wiedemann-Franz law (second line in Eq.~(13)). It is also striking that $\alpha_{yxx}, l_{yxx}, \sigma_{yxx}$ here follow the similar relations as found for the transport coefficients of the BCD induced nonlinear anomalous thermoelectric effects restricted by time reversal symmetry \cite{zeng2019wiedemannfranz}. Rather than appearing as a more conventional deviation from the conventional Wiedemann-Franz law and Mott relation, the introduced derivative in the Wiedemann-Franz law and the removed derivative in the Mott relation predicted here (Eq.~(13)) are attributed to the chiral anomaly related intrinsic nonlinearity.

In this section, we have derived the equations for the chiral anomaly induced nonlinear planar Nernst and thermal Hall effects in the presence of a coplanar thermal gradient and magnetic field ($\bm{B \cdot \nabla}T \neq 0$), and have also obtained their relations in a low temperature regime. In what follows, we apply these equations to the WSMs using a linearized Weyl Hamiltonian as well as a lattice Weyl Hamiltonian. 

\section{Nonlinear planar thermoelectric transport in Weyl Semimetals}

\subsection{linearized low-energy model for single Weyl cone}
\begin{figure}[tp!]
	\begin{center}
		\includegraphics[width=0.46\textwidth]{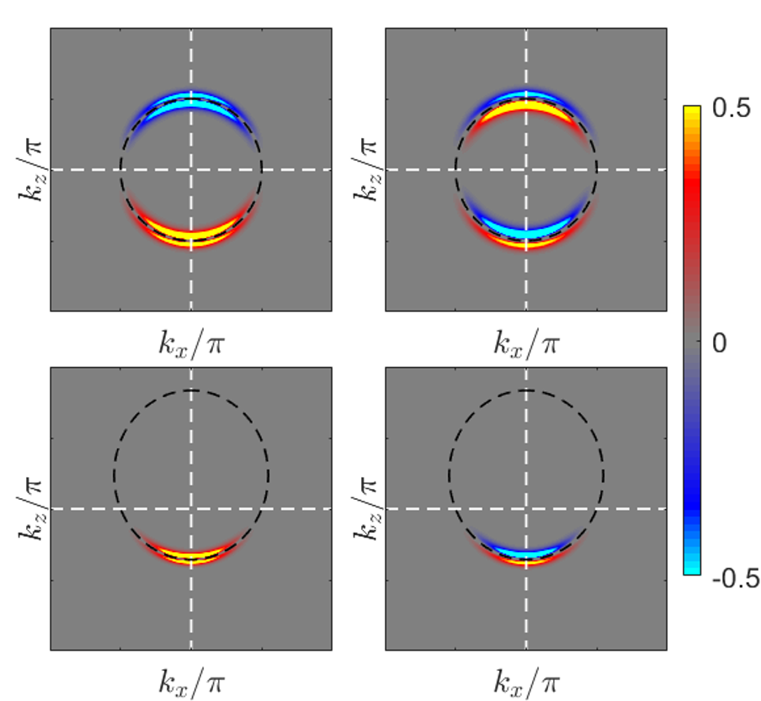}
		\llap{\parbox[b]{163mm}{\large\textbf{(a)}\\\rule{0ex}{70mm}}}
		\llap{\parbox[b]{90mm}{\large\textbf{(b)}\\\rule{0ex}{70mm}}}
		\llap{\parbox[b]{166mm}{\large\textbf{(c)}\\\rule{0ex}{35mm}}}
		\llap{\parbox[b]{93mm}{\large\textbf{(d)}\\\rule{0ex}{35mm}}}
	\end{center}
	\caption{(Color online) Modulated Berry curvature (a) \textcolor{black}{$\widetilde{\bm{\Omega}}^{\alpha}_{\bm{k}}$} and (b) \textcolor{black}{$\widetilde{\bm{\Omega}}^{l}_{\bm{k}}$} projected in the $k_x-k_z$ plane for non-tilted Weyl cone described by Hamiltonian given in Eq.~(14) respectively. Panel (c) and (d) display the similar projections as in (a), (b) but for a Weyl cone with tilt strength $R_s =0.4$. The black dashed lines indicate the zero-temperature Fermi surface at $\mu=0.2~v_F \hbar$ for the Weyl cone and the colors indicate the magnitude of the modulated Berry curvature nearby the Fermi surface, \textcolor{black}{which is now normalized by their corresponding maximum in each panel}. The other parameters used here are $n=1, s=1, v_F=1 eV$, $k_{x,z} \in [-0.16\pi,0.16\pi]$, a finite temperature $T=50K$ (i.e., $\beta=230 (eV)^{-1}$) is applied here.
	} \label{fig:fig1}
\end{figure}
\begin{figure}[htp!]
	\begin{center}
	\includegraphics[width=0.46\textwidth]{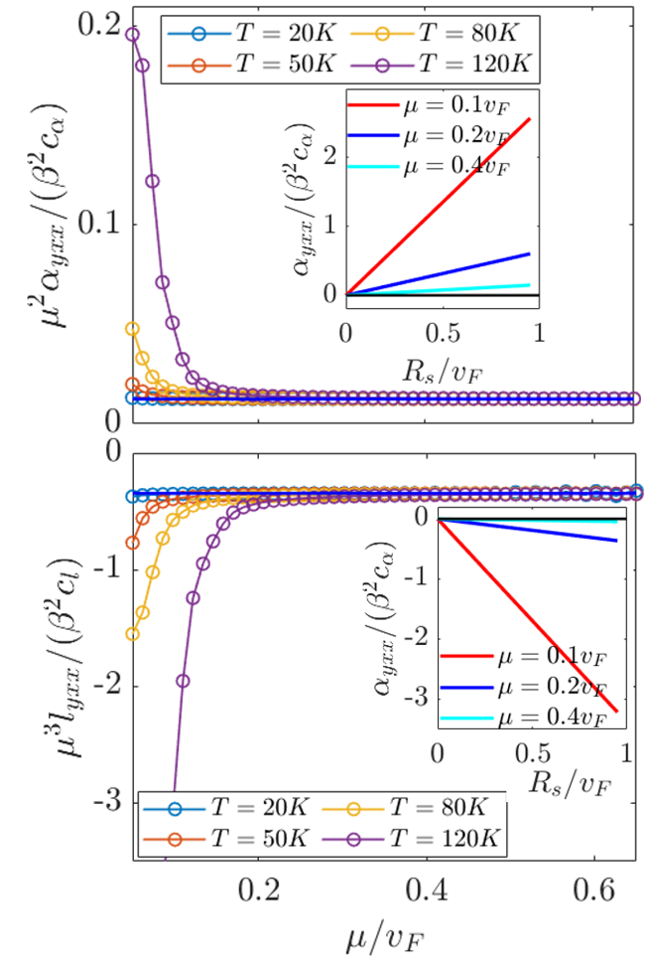}
		\llap{\parbox[b]{155mm}{\large\textbf{(a)}\\\rule{0ex}{118mm}}}
	\llap{\parbox[b]{156mm}{\large\textbf{(b)}\\\rule{0ex}{60mm}}}
	\end{center}
	\vspace{-5mm}
	\caption{(Color online) Nonlinear planar Nernst coefficient $\alpha_{yxx}$ and thermal Hall coefficient $l_{yxx}$ for a tilted Weyl cone ($R_s=0.5v_F$) plotted as a function of chemical potential $\mu$ in panel (a) and (b) respectively. The circled lines represent for the numerical results at different temperatures based on Eq.~(10) while the horizontal solid blue lines indicate the analytical (as well as zero-temperature) predictions for $\alpha_{yxx}, l_{yxx}$ using Eq.~(13). Note that, for the sake of a better demonstration of the chemical potential dependencies, we have multiplied $\mu^2, \mu^3$ to $\alpha_{yxx}, l_{yxx}$ respectively. The inset shows that the nonlinear planar transport coefficients are linearly dependent on the tilt strength of the Weyl cones. The other parameters used here are the same as that of Fig.~1. }
	\label{fig:fig2}	\vspace{-5mm}
\end{figure}
The linearized low-energy effective Weyl Hamiltonian for a single Weyl node can be given as \cite{Weylmodel_2016jetp}, 
\begin{equation}
    \begin{split}
    H= \hbar v_F R_s (k_z -s Q) \sigma_0 +s \hbar v_F(\bm{k}-s Q \bm{\hat{z}})\cdot \bm{\sigma}
    \end{split}
\end{equation}
where $s=\pm 1$ indicates the chirality of the Weyl node, $R_s, 2Q$ respectively represent the tilting strength and the separation of the Weyl nodes in momentum space
, $\sigma_0$ and $\bm{\sigma}$ are $2 \times 2$ Pauli matrices. According to the degree of tilt around the Weyl node, we have type-I ($|R_s| <1$) and type-II ($|R_s|>1$) Weyl cones. The eigenenergy for Eq.~(14) is given by $\e^s_{\bm{k}}=  \hbar v_F (R_s \tilde{k}_z \pm \sqrt{k^2_{\perp}+\Tilde{k}^2_{z}})$ with $k_{\perp}=\sqrt{k^2_x+k^2_y}, \tilde{k}_z=k_z -sQ$, and the Berry curvature is calculated as
\begin{equation}
    \begin{split}
    \bm{\Omega_k}^s = \mp s \frac{\bm{k}-s Q\hat{\bm{z}}}{2|\bm{k}-s Q\hat{\bm{z}}|^3}
    \end{split}
\end{equation}
which is impervious to the tilt parameter. Regarding the Weyl node separation $2Q$ in momentum space, it only shifts the distribution of the Berry curvature in momentum space without any modification and won't affect any physical property in the single Weyl node model. Hence, we can hereafter set $Q=0$ for the following analysis without affecting any results in this current work. 
With the help of eigenenergy and Berry curvature, we can calculate the corresponding nonlinear planar Nernst coefficient $\alpha_{yxx}$ and thermal Hall coefficient $l_{yxx}$ for the given Weyl Hamiltonian in Eq.~(14). Note that, in the configuration of a coplanar magnetic field and thermal gradient lying in $x-y$ plane, the applied magnetic field is perpendicular to the direction of the tilt strength as well the momentum separation of the Weyl cones. 



In Fig.~1, we show the $k_x$-$k_z$ plane projection of the modulated Berry curvature $\widetilde{\bm{\Omega}}_{\bm{k},\alpha(l)}$ without/with tilt. As shown in Fig.~1, a finite tilt can not only shift the Fermi surface (black dashed lines) along $k_z$-axis, it also results in an asymmetric distribution of the modulated Berry curvature (indicated by the colors) with respect to the Fermi surface. Because of this, a finite tilt strength leads to a non-zero net contribution on the Fermi surface and thus, is indispensable for the nonlinear planar transport coefficients to be non-zero. Note that, to better track the tilt induced changes of the distribution of the modulated Berry curvature $\widetilde{\bm{\Omega}}_{\bm{k},\alpha(l)}$ nearby the Fermi surface, they are now normalized by their own maximum value in each panel of Fig.~1. 

Based on Eq.~(12) and (14), we can now straightforwardly calculate the zero-temperature chiral anomaly induced nonlinear Hall effect (CNHE) coefficient $\sigma^0_{yxx}$. We found $\sigma^0_{yxx}= c^{\prime}_{\sigma} R_s/\mu^2$ ($c^{\prime}_{\sigma}$, the modified constant factor), consistent with what has been obtained in Ref. \cite{chiralanomaly_2020Li}. Note that, $\sigma^0_{yxx}$ is linearly tilt-dependent, so are the coefficients $\alpha_{yxx}, l_{yxx}$ according to Eq.~(13). It will be interesting to numerically check the chemical potential dependency of $\alpha_{yxx}, l_{yxx}$ for a linearized Weyl Hamiltonian. In Fig.~2, we plot the nonlinear planar thermoelectric transport coefficients $\alpha_{yxx}, l_{yxx}$ as a function of the chemical potential in panel (a), (b) respectively. The circled lines are numerical results at different temperatures based on Eq.~(10), while the horizontal blue solid lines are based on the analytical results in Eq.~(13) (along with $\sigma^0_{yxx}=c^{\prime}_{\sigma} R_s/\mu^2$). At chemical potentials away from the Weyl node ($\mu>0.2v_F$), both $\mu^2 \alpha_{yxx}$ (top panel) and $\mu^3 l_{yxx}$ (bottom panel) are convergently proportional to a constant at different temperatures, indicating the chemical potential dependency of $\alpha_{yxx} \propto \mu^{-2}, l_{yxx} \propto \mu^{-3}$ respectively. The insets in Fig.~(2) show us a linear dependence on the tilt strength for $\alpha_{yxx}$ and $l_{yxx}$ with different chemical potentials (indicated by the colors) in the top and bottom panels respectively, consistent with $\sigma^0_{yxx}\propto R_s$ as we analytically obtained earlier.

\begin{figure}[htp!]
	\begin{center}
	\includegraphics[width=0.46\textwidth]{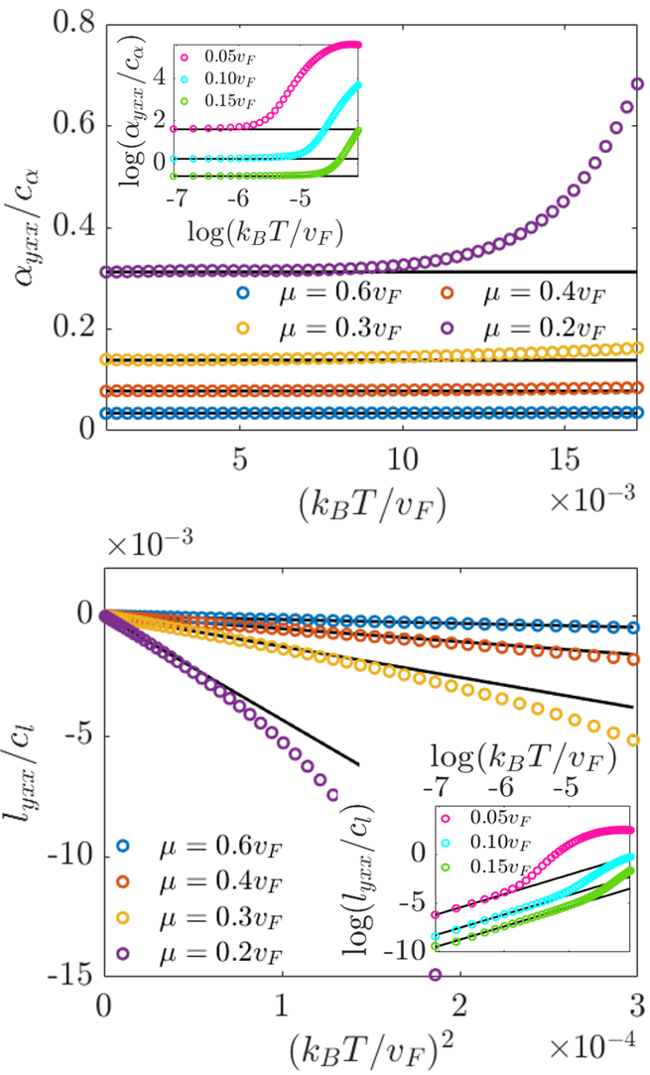}
		\llap{\parbox[b]{162mm}{\large\textbf{(a)}\\\rule{0ex}{132mm}}}
	\llap{\parbox[b]{163mm}{\large\textbf{(b)}\\\rule{0ex}{63mm}}}
	\end{center}
	\vspace{-5mm}
	\caption{(Color online) Nonlinear planar Nernst coefficient $\alpha_{yxx}$ and thermal Hall coefficient $l_{yxx}$ for a tilted Weyl cone ($R_s=0.5v_F$) plotted as a function of temperature in panel (a) (vs $(k_B T)$) and (b) (vs $(k_B T)^2$) respectively. The circles represent for the numerical results at different chemical potentials based on Eq.~(10), while the corresponding black solid lines indicate the analytical predictions by Eq.~(13). The zero slops and finite constant slops in panel (a) and (b) respectively imply $\alpha_{yxx} \propto (T)^0$ and $l_{yxx} \propto (T)^2$. 
	The insets show the logarithm plot of $\alpha_{yxx}$ and $l_{yxx}$ with several lower chemical potentials in top and bottom panel, respectively.
	}
	\label{fig:fig3}
\end{figure}
We also numerically check the behavior of the nonlinear transport coefficients $\alpha_{yxx},l_{yxx}$ as a function of temperature, as shown in the top and bottom panel in Fig.~(3) respectively. As predicted by the analytical expression approximated in low temperature regime given in Eq.~(13), it is evidently shown in Fig.~3 that $\alpha_{yxx} \propto T^0$ (top panel) while $l_{yxx} \propto T^2$ (bottom panel) at chemical potentials away from the Weyl node ($\mu >0.2v_F$). The black solid lines corresponding to each chemical potential are based on the analytical expression in Eq.~(13). The logarithm plot for $\alpha_{yxx}(T), l_{yxx}(T)$ with some lower chemical potentials are also presented in the inset of top and bottom panel, respectively.
\textcolor{black}{The deviations appearing at relatively higher temperatures or lower chemical potentials in Fig.~2 and Fig.~3 can be attributed to the omitted higher order terms in Eq.~(13). When dealing with chemical potential closer to the Weyl cones, we must count in the contributions from the higher order terms for $\sigma_{yxx}, \alpha_{yxx}$ and $l_{yxx}$ to obtain the more correct descriptions. }

So far, we have numerically calculated the nonlinear planar Nernst and thermal Hall coefficients (Eq.~(10)) as well as checked their nonlinear analog of the Wiedemann-Franz law and the Mott formula (Eq.~(13)), using a simple low-energy effective Weyl Hamiltonian (Eq.~(14)). Within this context, 
to get the finite (nonzero) chiral anomaly induced nonlinear planar response functions, the Weyl cones in each pair are required to be tilted along non-opposite axial directions. This is indicated by the Eq.~(13) along with the analytical results for CNHE with $\sigma^0_{yxx}=c^{\prime}_{\sigma}R_s/\mu^{2}$ \cite{chiralanomaly_2020Li}. 
In the next section via the lattice model, we find 
that with finite chiral chemical potential, the Weyl node pairs tilted in opposite direction with respect to each other can still lead to a non-zero value for nonlinear planar Nernst and thermal Hall effects. Interestingly, we also find that finite nonlinear response functions can even exist when the Weyl cones are not tilted at all due to the lattice regularization. This result is contrary to what was found for chiral anomaly induced non-linear planar Hall effect based on low energy effective model in Ref. \cite{chiralanomaly_2020Li}.

\subsection{Lattice model for a pair of oppositely titled Weyl cones}

As have been pointed out in the recent theoretical studies on 3D topological Dirac and Weyl semimetals \cite{Girish_2016_THE, AHE_weyl2020czeng}, there are some significant features that can not be captured by the low-energy linearized model Hamiltonian with respect to the realistic materials, especially for that originate from the Fermi surface contributions. Therefore, it will be necessary to look at the chiral anomaly related nonlinear planar response functions using the lattice model.
\begin{figure}[tp!]
	\begin{center}
	\includegraphics[width=0.46\textwidth]{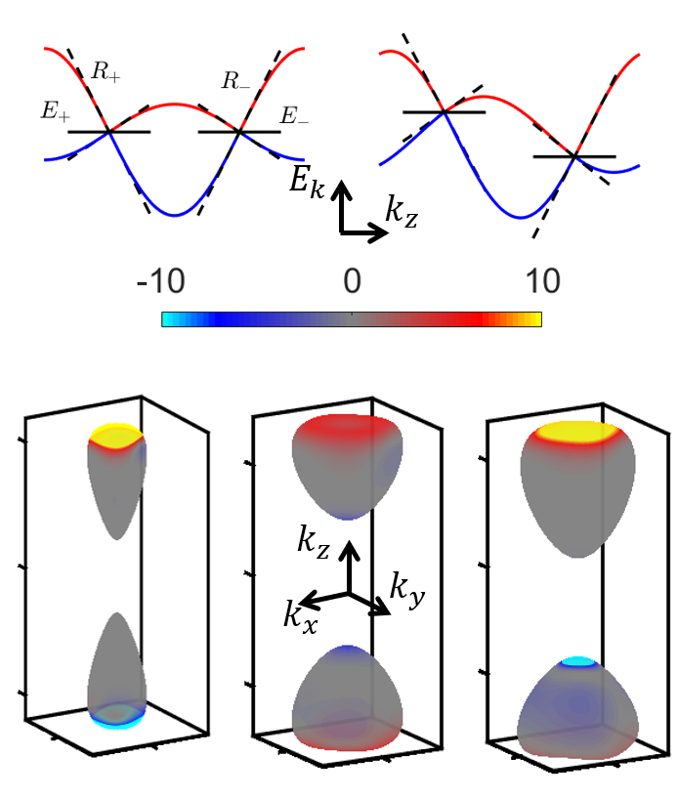}
		\llap{\parbox[b]{83 mm}{\large\textbf{(a)}\\\rule{0ex}{91 mm}}}
	\llap{\parbox[b]{161mm}{\large\textbf{(b)}\\\rule{0ex}{55mm}}}
	\end{center}
	\vspace{-5mm}
	\caption{(Color online) (a) Schematic band structures for the lattice Hamiltonian in Eq.~(16). The tilt strength and the intrinsic energy for each Weyl node are described by $R_{\pm}, E_{\pm}$ respectively. The left panel in (a) represent a pair of Weyl nodes with opposite tilt strength ($R_{+}=R_{-}$) at same energies ($E_{+}=E_{-}$), while the right panel in (a) shows a pair of Weyl nodes with opposite tilt at different energies $E_{+}=-E_{1}=\delta E$. (b) The distribution of modulated Berry curvature $\widetilde{\bm{\Omega}}_{\bm{k},\sigma}$ on Fermi surfaces of lattice Weyl Hamiltonian (Eq.~(16)) with \textit{left: }$t_1 =0.5 t, \delta E=0$, \textit{middle: }$t_1=0, \delta E=1.6t$ and \textit{right: }$t_1=0.6 t, \delta E=1.6t$ at Fermi energy $\mu=0.9t, 0, 0.4t$, respectively. Shown by the middle and right panel in (b) respectively, in the presence of finite $\delta E$, the net contribution of the modulated Berry curvature on the Fermi surface is non-zero even though the Weyl cones are not tilted or oppositely tilted. }
	\label{fig:fig4}
\end{figure}
\begin{figure}[tp]
	\begin{center}
	\includegraphics[width=0.47\textwidth]{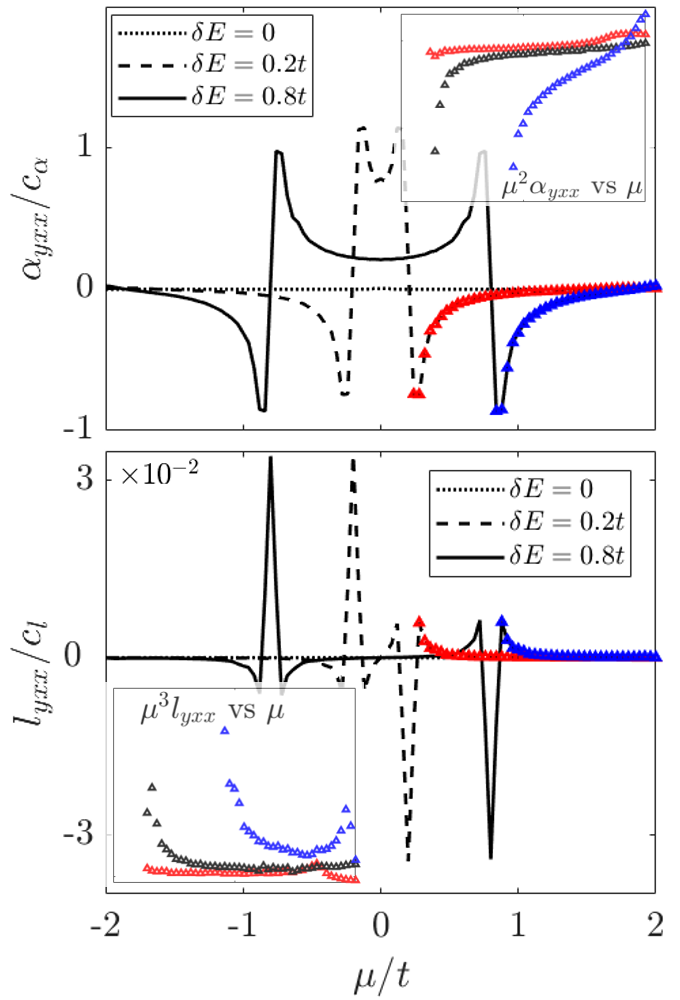}
		\llap{\parbox[b]{160 mm}{\large\textbf{(a)}\\\rule{0ex}{118 mm}}}
	\llap{\parbox[b]{161mm}{\large\textbf{(b)}\\\rule{0ex}{62mm}}}
	\end{center}
	\vspace{-5mm}
	\caption{(Color online) Similar plots as Fig.~2 for lattice Weyl Hamiltonian in Eq.~(16) describing a pair of nontilted Weyl cones ($t_1=0$) with different chiral chemical potentials $\delta E=0, 0.2t, 0.8t$. Both $\alpha_{yxx}$ (top) and $l_{yxx}$ (bottom) show an evident peak when chemical potential goes cross the Weyl nodes, and they identically vanish when $\delta E=0$ (dotted lines). The red and blue triangles in panel (a) (panel (b)) represent $\alpha_{yxx}$ ( $l_{yxx}$) at chemical potentials above the top Weyl node lying at $E=0.2t, 0.8t$ respectively. The corresponding inset in the top and bottom panel showing roughly horizontal dependencies for Weyl nodes with relatively small $\delta E$ (blue and black triangles), reveals the chemical potential dependence $\alpha_{yxx}\propto \mu^{-2}$ and $l_{yxx} \propto \mu^{-3}$ respectively. Note that, comparing to the red triangle data points in the inset plots, the black triangle data correspond to the Weyl nodes with $\delta E=0.2t$ and tilt strength $t_1 =0.24t$. The other parameters used here are same as that in Fig.~4. }
	\label{fig:fig5}
\end{figure}

In this section, we work on a lattice model describing a pair of tunable Weyl nodes to prob the nonlinear planar thermoelectric effects induced by the thermal chiral anomaly, and the Hamiltonian is given as the following \cite{hybrid_2016chen},
\begin{equation}
    \begin{split}
        H(\bm{k}) &=N_0((\bm{k}))\sigma_0 +\bm{N(k)\cdot \sigma},\\
        N_{x}(\bm{k}) &=2 t^{\prime}_x\sin{k_x},  N_{y}(\bm{k}) = 2t^{\prime}_y \sin{k_y},\\
         N_{z}(\bm{k}) &= (m-2t_x \cos{k_x}-2t_y\cos{k_y}-2t_z \cos{k_z}),\\
         N_0{(\bm{k})}&=2t_1 \cos{(\phi_1 - k_z)} +2t_2\cos{(\phi_2-2k_z)}. 
    \end{split}
\end{equation}
It is straightforward to check that the above Hamiltonian is both time reversal symmetry broken and inversion symmetry broken, i.e., $\mathcal{T}H(\bm{k})\mathcal{T^{\dagger}}\neq H(-\bm{k})$, and  $\mathcal{P}H(\bm{k})\mathcal{P^{\dagger}} \neq H(\bm{-k})$ with $\mathcal{P}=\sigma_x$ and $\mathcal{T}=\mathcal{K}$ with $\mathcal{K}$ the anti-Hermitian complex conjugation operator. A pair of Weyl nodes can be found located at $(0,0,\pm k_0)$ in momentum space with $k_0$ satisfying $2t_z \cos{k_0}=m-2t_x-2t_y$ for Eq.~(16). In effect, the term $N_{0}(\bm{k})$ can generally lift the inversion like and particle-hole like symmetries of $\bm{N(k)\cdot \sigma}$ by modulating the energies and tilt of the Weyl nodes \cite{hybrid_2016chen}. By tuning the tilting term $N_0(\bm{k})$, or explicitly the parameters $t_1, t_2$, the above lattice Hamiltonian can effectively describe Weyl systems with different types of Weyl nodes. For example, both two Weyl cones belong to type-I or type-II Weyl cone, or one of the Weyl cones belongs to type-I while the other belongs to type-II, which corresponds the so-called hybrid Weyl semimetal \cite{hybrid_2016chen,hybrid_exp2018Shi}. It is obvious that the chiral anomaly induced nonlinear planar response functions (i.e., $\alpha_{yxx},l_{yxx}$) will be finite for a hybrid Weyl system of a pair of Weyl nodes with different tilt strengths, i.e., $R_{+}\neq -R_{-}$. To get a clear vision of the effect of band regularization on the chiral anomaly induced nonlinear planar effects, we center us upon the Weyl systems with a pair of type-I Weyl cones tilted in opposite direction, by setting $t_2=0$ in Eq.~(16) hereafter in this paper. 

\textcolor{black}{Without loss of generality, we can choose the parameters $m, t_x, t_y$ such that $k_0 =\pi/2$. 
Thus the term $N_0(\bm{k})$ approaching to the Weyl nodes at $(0, 0, \pm k_0)$ can be rewritten as 
\begin{equation}
    \begin{split}
        N_0({\bm{k^{\prime}}})= s\delta E + 2 t_{1}s \sqrt{1-(\delta E/2t_1)^2} k^{\prime}_{z}
    \end{split}
\end{equation}
provided the phase factor $\phi_1  =\pi+\sin^{-1}{[\delta E/(2t_1)]}(t_1 \neq 0)$. Here $s=\pm 1$ indicating the opposite chiralty of the Weyl nodes, and $\bm{k}^{\prime}$ is the momentum measured from the position of Weyl nodes. It is obvious that the hybrid Weyl Hamiltonian in Eq.~(16) now  depicts a pair of Weyl nodes with effective tilt strength $\widetilde{R}_s=2t_1 s\sqrt{1-(\delta E/(2t_1))^2} $ and chiral chemical potential $2\delta E$, respectively.}
\begin{figure}[htp!]
	\begin{center}
	\includegraphics[width=0.47\textwidth]{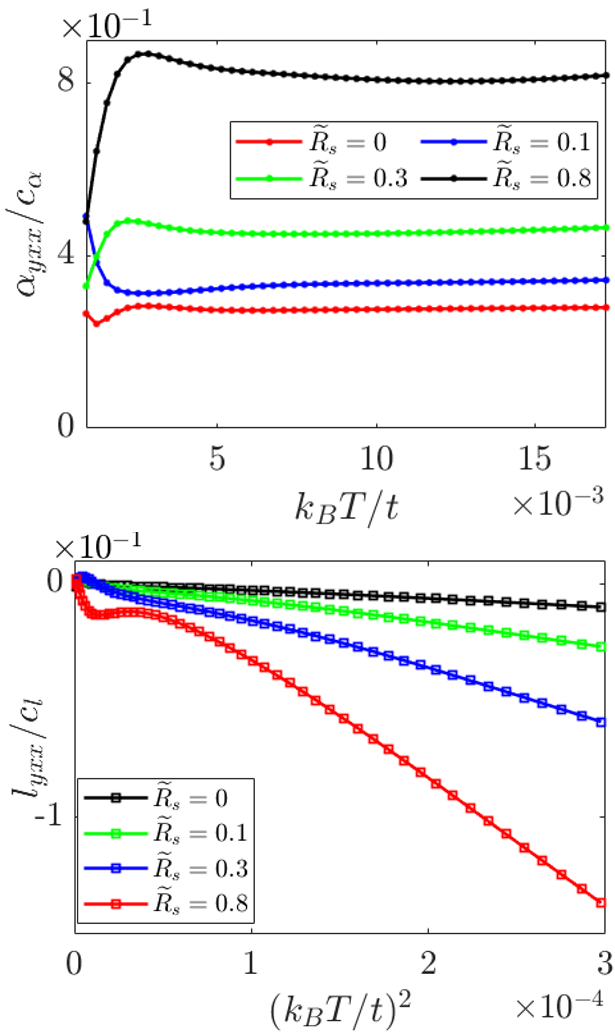}
		\llap{\parbox[b]{163 mm}{\large\textbf{(a)}\\\rule{0ex}{130 mm}}}
	\llap{\parbox[b]{162mm}{\large\textbf{(b)}\\\rule{0ex}{70mm}}}
	\end{center}
	\vspace{-5mm}
	\caption{(Color online) Similar plots as Fig.~3 for lattice Weyl Hamiltonian given in Eq.~(16) describing a pair of oppositely tilted Weyl nodes with finite chiral chemical potential $\delta E=0.8t$. The effective tilt strength $\widetilde{R}_s =0, 0.1, 0.3,0.8$ respectively corresponds to parameter $t_1 =0.4t, 0.413t, 0.5t$ and $0.9t$. The temperature ranges from $10K$ to $200K$ for both panels. As shown in the top panel, $\alpha_{yxx}$ remains finite approaching to zero temperature and tend to be a constant with increasing temperatures, i.e., $\alpha_{yxx}\propto T^0$. Contrarily, $l_{yxx}$ in the bottom panel vanishes at zero temperature and linearly depend on $(k_BT)^2$, i.e., $l_{yxx} \propto T^2$. The chemical potential here is $\mu=-0.4t$, and the other parameters used here are the same as that in Fig.~4.}\label{fig:fig6} 	\vspace{-3mm}
\end{figure}
As schematically shown in Fig.~4 (a), the two Weyl cones are now oppositely tilted with $R_{+}=-R_{-}$. In the case of zero chiral chemical potential i.e., $E_{+}=E_{-}=0$ (left panel in Fig.~4 (a)), the contribution of the modulated Berry curvature (indicated by the colors) on the Fermi surfaces around each Weyl node are well cancelled with each other, as shown by the left panel in Fig.~4 (b). When considering a finite chiral chemical potential i.e., $E_{+}\neq E_{-}$ (shown by right panel, Fig.~4 (a)), the net value of the modulated Berry curvature on the Fermi surface can be non-zero either with zero tilt (middle panel, Fig.~4 (b)) or opposite tilt (right panel, Fig.~4 (b)) for the Weyl cones. Evidently, for a 3D lattice Weyl system, the unequal energies of the Weyl nodes (or finite chiral chemical potential) naturally introduce an asymmetric Fermi surface with the help of the lattice regularization. \textcolor{black}{This in turn gives rise to nonzero net value for the nonlinear transport coefficients [Eq.~(10)], which are purely Fermi surface quantities.}
We want to point out that, this feature depicted by the right panel with $R_{+}=-R_{-}\neq 0$ (but not the middle panel, $R_{+}=R_{-}=0$) in Fig.~4 (b) can actually be revealed within the regime of linearized low-energy Hamiltonian model by considering an effective chemical potential $\widetilde{\mu}=\mu+E_{s}$. For instance, the aforementioned zero-temperature NPHE coefficient $\sigma^0_{yxx}$ becomes finite ($\sum_{s} c^{\prime}_{\sigma} R_s/\widetilde{\mu}^2 \neq 0$) assuming $R_{+}=-R_{-}\neq 0$ and $E_{+}\neq E_{-}$. 


Using the equations (Eqs.~(7), (9), (10)) formalized by semiclassical Boltzmann transport approach in Sec. II, we can similarly obtain the different parameters' dependencies for the chiral anomaly induced nonlinear planar response functions for the lattice Hamiltonian.  

In Fig.~5, we plot the nonlinear planar transport coefficient $\alpha_{yxx}, l_{yxx}$ for the lattice Weyl Hamiltonian in Eq.~(16) with zero tilt strength ($t_1=0, \widetilde{R}_s =0$) as a function varying with chemical potentials. Due to the divergently large Berry curvature (so does the modulated Berry curvature) around the Weyl nodes, both $\alpha_{yxx}$ and $l_{yxx}$ show an evident enhancement in magnitude whenever the Fermi energy corsses the Weyl nodes at $\mu=\pm \delta E$, as shown in Fig.~5. In the non-tilted Weyl systems, a finite chiral chemical potential tends to be determinant for $\alpha_{yxx}, l_{yxx}$ to be non-zero, as can be seen in Fig.~5. When the chemical potential lies relatively away from the Weyl nodes, the relations $\alpha_{yxx} \propto \mu^{-2}$ while $l_{yxx} \propto \mu^{-3}$ found for the single Weyl node earlier (Fig.~2) still hold for the lattice Weyl system, as expected. For instance, for the data points within the chemical potential range implied by the triangles in Fig.~5, the above relationships can be directly revealed by the inset plot of $\mu^2 \alpha_{yxx}$ and $\mu^3 l_{yxxx}$ versus $\mu$ in the top and bottom panel in Fig.~5 respectively. 

We can also check how the temperature affects the nonlinear planar Nernst and thermal Hall coefficients for the lattice Hamiltonian model of the 3D WSMs. By plotting $\alpha_{yxx}, l_{yxx}$ as a function of $(k_B T), (k_B T)^2$, we get the temperature dependencies $\alpha_{yxx} \propto T^0$ and $l_{yxx} \propto T^2$ as respectively displayed in Fig.~6 (a), (b). Note that, the coefficient $\alpha_{yxx}, l_{yxx}$ for the non-tilted Weyl node pair ($\widetilde{R}_s=0$) with finite chiral chemical potential ($\delta E=0.8t$) are also non-zero (consistent with Fig.~5), and they follow the similar temperature dependencies, as shown by the black circled and black squared line in the top panel and bottom panel in Fig.~6 respectively. The nonlinear planar Nernst coefficient $\alpha_{yxx}$ remains finite even at the zero temperature limit ($T\rightarrow 0$) while the nonlinear planar thermal Hall coefficient $l_{yxx}$ vanishes when temperature approaches to zero. Consequently, these $\mu$-dependencies and $T$-dependencies numerically found here (in Fig.~5, 6) for the chiral anomaly induced nonlinear planar transport coefficients via the lattice Weyl Hamiltonian model (Eq.~(16)), also agree well with the fundamental relations derived as the nonlinear analog of the Wiedemann-Franz law and Mott formula given in Eq.~(13). We want to mention that, the data would be smoother if the resolutions for the grid points in momentum space are refined in the numerical calculations.  

In this section, we have focused on the lattice model describing a pair of Weyl cones tilted in opposite direction with respect to each other as well as located at different energies. We find that the location of the Weyl nodes at different energies naturally introduce an asymmetric Fermi surface and therefore, rendering the net contributions to the nonlinear planar Nernst and thermal Hall coefficients from the two oppositely tilted Weyl cones to be nonzero. The nontrivial results we found using a lattice Weyl Hamiltonian in this section indeed restore the possibility to testify the nonlinear planar Nernst and thermal Hall effects induced by the thermal chiral anomaly and prob their related parameter dependencies in the realistic 3D WSMs in experiment.   
\vspace{5mm}

\section{Summary and conclusion}
By utilizing the semiclassical Boltzmann transport approach, we investigate the second order (in terms of the thermal gradient $\bm{\nabla}T$) nonlinear planar Nernst and thermal Hall effects (Eq.~(11)) in 3D WSMs, which can be viewed as the manifestations of chiral anomaly with $\bm{B \cdot \nabla}T \neq 0$. In the low temperature regime, we also derive the fundamental relations among the chiral anomaly induced nonlinear planar transport coefficients via the Sommerfeld expansions. Interestingly, the chiral anomaly induced nonlinear transport phenomena violate the well-known Wiedemann-Franz law and Mott relation derived in the linear response regime \cite{Lorenz_law,Ashcroft_my}. Instead, the nonlinear analog of these two celebrated equations found in the present work (Eq.~(13)) are remarkably consistent with that predicted for the BCD induced nonlinear transport phenomena in Ref. \cite{zeng2019wiedemannfranz}. 

We have numerically checked our results using both the low energy effective Weyl Hamiltonian as well as the lattice Weyl Hamiltonian. Compared to the simple linearzied Weyl Hamiltonian models, the lattice models generally acquire band structures more closely related to the realistic WSMs, because of the natural band regularization and irreproducible overlap regions between the Weyl nodes. We find that, to get a non-zero nonlinear planar response induced by chiral anomaly, the configurations of the pairs of Weyl nodes of opposite chiralities are not limited to those tilted in the same directions, consistent to what has been found in a recent work for the nonlinear planar Hall effect \cite{chiralanomaly_2020Li}. In contrast to Ref. \cite{chiralanomaly_2020Li}, we show that, pairs of non-tilted or oppositely tilted Weyl cones can also
support nonzero chiral anomaly induced NPNE and NPTHE in Weyl materials. We have also given concrete prescriptions for experimentally distinguishing the effects discussed in the present work from other similar linear and non-linear effects that may also arise in this scenario. The novel non-linear effects in WSMs induced by chiral anomaly in the presence of $\bm{E=0}$ and $\bm{B \cdot \nabla}T \neq 0$ and the fundamental relations among them remarkably different from the conventional Wiedemann-Franz law and Mott relations can be checked experimentally in realistic WSMs. 

\section{Acknowledgments} 
C. Z. acknowledges support from the National Key R\&D Program of China (Grant
No. 2020YFA0308800). S. N. acknowledges the National Science Foundation Grant No. DMR-1853048. S. T. thanks the ARO Grant No. W911NF-16-1-0182 and Grant No. NSF 2014157 for support.
\appendix*
\section{Contributions of intra-node and inter-node scatterings to the nonlinear thermoelectric effects}
In this appendix, we investigate the contributions of intra-node and inter-node scatterings to the chiral anomaly induced nonlinear Nernst and thermal Hall effects. 
With the relaxation time approximation, the collision term for WSMs in Eq.~(1) in the main text can be described by the following equation, where both the intra-node ($\tau_a$) and inter-node ($\tau_v$) relaxation times are involved, 
\begin{equation}
    \mathcal{I}_{coll}(f^{s}_{\bm{r,k}}) = -\frac{f^{s}_{\bm{r,k}}-f^{s}_{eq}}{\tau_a} -\frac{f^{s}_{\bm{r,k}}-f_{eq}}{\tau_{v}} 
\end{equation}
Here $s=\pm 1$ represent the chirality of the Weyl node, $f^{s}_{eq}$ and $f_{eq}$ indicate the local (with respect to the Weyl node $s$) and global equilibrium state achieved via intra-node and inter-node scattering processes, respectively. To proceed, instead of the more detailed expansions as given in Eq.~(4) in the main text, here we apply the following relations, 
\begin{equation}
    \begin{split}
        f^{s}_{\bm{r,k}} &= f_{eq} +[-\partial_{\e_{\bm{k}}} f_{eq}(\bm{r,k})] \delta \e^s_{\bm{k}},\\
    f^{s}_{eq} &= f_{eq} +[-\partial_{\e_{\bm{k}}} f_{eq}(\bm{r,k})] \delta \mu^{s},\\
    \end{split}
\end{equation}
where $\delta \e^s_{\bm{k}}$ is the energy difference caused by the external fields, and $\delta \mu^s =\mu^s -\mu$ implies the chemical potential imbalance for Weyl node $s$. Consequently, substituting the above equations into the steady-state Boltzmann transport equation in Eq.~(3), we obtain, 
\begin{equation}
    \begin{split}
         \big[-e\bm{v}(\bm{B\cdot \Omega})/\hbar + \frac{e}{\hbar} (\bm{ v\cdot \Omega})\bm{B}\big] \frac{\e_{\bm{k}}-\mu}{T}\bm{\nabla}T  =& \\
         -\frac{\delta \e^s_{\bm{k}}-\delta \mu^s}{\tau_a} -\frac{\delta \e^s_{\bm{k}}}{\tau_{v}} &
    \end{split}
\end{equation}
Here only the terms linear in magnetic field are shown on the left hand side, and the factor $[-\partial_{\e_{\bm{k}}} f_{eq}(\bm{k})]$ is canceled for both sides. 

It is not hard to get aware that,  $\langle f^s_{\bm{r, k}}\rangle_{s} =f^s_{eq}$ and $\langle \delta \e^s_{\bm{k}} \rangle_{s} =\delta \mu^s$, where $\langle\dots \rangle_{s}$ denotes the average over all the possible electron states with respect to Weyl node $s$ [see Eq.~(7) in Ref.~\cite{Deng_qc_2019prl}]. Taking the average for the both sides of Eq.~(A.3), the chemical potential imbalance is found as
\begin{equation}
        \delta \mu^s = \tau_{v} c_0 (\bm{\nabla} T \bm{\cdot B}) \langle \bm{\Omega} \cdot\bm{v}\frac{\e_{\bm{k}}-\mu}{T}\rangle_{s} 
\end{equation}
where $c_0$ is constant factor. Here we only focus on the general format for $\delta \mu^s$ hence the detailed expression for the average in Eq.~(A.4) will not be discussed. In general, we have $1/\tau_{v} \ll 1/\tau_a$ for WSMs. Thus it is physically allowed to make the approximation $\delta \e^s_{\bm{k}}/\tau_v \approx \delta \mu^s/\tau_v$ for the second term on the right hand side in Eq.~(A.3). As a result, the distribution function deviation $\delta f_{\bm{k}}= f^{s}_{\bm{r,k}} - f_{eq} $ in presence of thermal gradient can be obtained based on Eqs.~(A.2-A.4), written as, 
\begin{equation}
    \begin{split}
        \delta f_{\bm{k}} &=[-\partial_{\e_{\bm{k}}} f_{eq}(\bm{k})] \big[ \delta \mu^s  -\tau_a\frac{\delta \mu^s}{\tau_v}   + \tau_a \dot{\bm{r}}_{eff} \big]
    \end{split}
\end{equation}
where $\dot{\bm{r}}_{eff}=\big[\bm{v}-e\bm{v}(\bm{B\cdot \Omega})/\hbar + \frac{e}{\hbar} (\bm{ v\cdot \Omega})\bm{B}\big]$ is the effective velocity under magnetic field. Obviously, the first term in the above equations comes from the chiral anomaly induced chiral chemical potential, which is generated by the thermal gradient and explicitly depends on the inter-node scattering time $\tau_v$. There is also a chiral anomaly related but $\tau_a$-dependent contribution, as well as a fully chiral anomaly-irrelevant contribution dependent on $\tau_a$. These results are consistent to what have been obtained in recent works~\cite{Deng_qc_2019prl,kamal_cas_2020prr}. 

As can be seen from Eqs.~(6-9) in the main text, it is the $\delta f_{\bm{k}}$ that lead to the nonlinear thermoelectric responses. For WSMs with $\tau_v\gg\tau_a$, we have $\delta f_{\bm{k}}\approx [-\partial_{\e_{\bm{k}}} f_{eq}(\bm{k})] \delta \mu^s$, i.e., the chiral anomaly induced chiral chemical potential makes the leading contribution to the nonlinear responses as given in Eq.~(7, 9) in the main text. 

\bibliography{my}


\end{document}